# Eighteen New Variable Stars in Cassiopeia and Variability Checking for NSV 364

**Riccardo Furgoni**

*Keyhole Observatory MPC K48, Via Fossamana 86, S. Giorgio di Mantova (MN), Italy; riccardo.furgoni@alice.it*

*and*

*AAMN Gorgo Astronomical Observatory MPC 434, S. Benedetto Po (MN), Italy*

**Abstract**   I report the discovery of eighteen new variable stars in Cassiopeia: eight pulsating (2MASS J00584964+5909260; GSC 03680-00667; GSC 03680-01320; GSC 03680-01488; 2MASS J02472793+6149024; GSC 04047-01118; GSC 04047-01418; GSC 04051-01789), six eclipsing (GSC 03680-00423; 2MASS J02443720+6143091; GSC 04047-00381; GSC 04047-00558; GSC 04051-02027; GSC 04051-02533), three rotating (ALS 6430; 2MASS J01020513+5912394; GSC 04051-01669), and one eruptive (GSC 04051-02483). The suspected variable star NSV 364 was checked for variability in six nights of observations and in the complete SuperWasp survey dataset; no variability was detected.

## 1. Introduction

A photometric campaign aimed at the discovery of new variable stars was carried out from the Keyhole Observatory in S. Giorgio di Mantova, Italy. Were observed two separate fields located in the constellation of Cassiopeia for a total of 21 nights, obtaining 2,329 images in the V passband with an exposure time of 2 minutes. Subsequently the light curves of all stars in the fields have been simply visually inspected in order to determine the candidate variables.

When possible the observations were combined with the SuperWASP and NSVS datasets in order to improve the precision in the determination of the period and the type of variability.

The large surface area of the sensor Kodak KAF8300 equipping the SBIG CCD ST8300m allowed us to observe a large portion of sky to a focal length suitable for areas crowded with stars.

This research shows how much work has to be done in the search for variable stars in crowded fields, especially in the area close to the Milky Way. An interesting concentration of variables was found not far from the star SAO 12415: in a radius of only 3.25' three new variables were discovered.

## 2. Instrumentation used

The data were obtained with a Celestron C8 Starbright, a Schmidt-Cassegrain optical configuration with aperture of 203 mm and central obstruction of 34%. The telescope was positioned at coordinates 45° 12' 33" N 10° 50' 20" E (WGS84) at the Keyhole Observatory, a roll-off roof structure managed by the author. The telescope was equipped with a focal reducer Baader Planetarium Alan Gee II able to bring the focal length from 2030 mm to 1413 mm in the optical train used. The focal ratio is also down to $f6.96$ from the original $f10$.

The pointing was maintained with a Syntha NEQ6 mount with software SYNSCAN 3.27, guided using a Baader Vario Finder telescope equipped with a Barlow lens capable of bringing the focal length of the system to 636 mm and focal ratio of $f10.5$.

The guide camera was a Magzero MZ-5 with Micron MT9M001 monochrome sensor equipped with an array of $1280 \times 1024$ pixels. The size of the pixels is $5.2\mu m \times 5.2\mu m$, for a resulting sampling of 1.68 arcsec/pixel.

The CCD camera was a SBIG ST8300m with monochrome sensor Kodak KAF8300 equipped with an array of $3352 \times 2532$ pixels. The pixels are provided with microlenses for improving the quantum efficiency, and the size of the pixels is $5.4\mu m \times 5.4\mu m$, for a resulting sampling of 0.80 arcsec/pixel.

The photometry in the Johnson V passband was performed with an Astrodon Photometrics Johnson V 50 mm round unmounted filter on a Starlight Xpress USB filterwheel.

The camera is equipped with a $1000 \times$ antiblooming: after exhaustive testing it has been verified that the zone of linear response is between 1,000 and 20,000 ADU, although up to 60,000 ADU the loss of linearity is less than 5%. The CCD is equipped with a single-stage Peltier cell $\Delta$-T$=35 \pm 0.1°$ C which allows cooling at a stationary temperature.

## 3. Data collection

The observed fields are centered, respectively, at coordinates (J2000) R.A. $01^h 01^m 12^s$, Dec. $+59° 05' 11"$ and R.A. $02^h 46^m 09^s$, Dec. $+61° 53' 12"$. For both the dimensions are $44' \times 33'$ with a position angle of $360°$.

The observations were performed with the CCD at a temperature of $-10°$ C in binning $1 \times 1$. The exposure time was 120 seconds with a delay of 1 second between the images and an average download time of 11 seconds per frame. The observations were conducted over twenty-one nights as presented in Table 1.

The CCD control program was CCDSOFT v5 (Software Bisque 2012). Once the images were obtained, calibration frames were taken for a total of 100 dark of 120 seconds at $-10°$ C, 200 darkflat of 2 seconds at $-10°$ C, and 50 flat of 2 seconds at $-10°$ C. The darkflats and darks were taken only during the first observing session and used for all other sessions. The flat were taken for each session as the position of the CCD camera could be varied slightly, as well as the focus point.

The calibration frames were combined with the method of the median and masterframes obtained were then used for the correction of the images taken. All images were then aligned and an astrometric reduction made to implement the astrometrical coordinate system WCS in the FITS header. These operations were conducted entirely through the use of software MAXIMDL v5.23 (Diffraction Limited 2012).

## 4. Finding comparison stars in the observed fields: the usefulness of the APASS data

Although the presence of an AAVSO comparison star sequence is a source of security in differential photometry, is not always possibile to have one for every observed field. In addition, since the APASS data were published (Henden *et al.* 2013) and are now available in Data Release 7, one can find a very accurate photometric reference for almost 97 percent of the sky.

I noticed that the APASS magnitudes are highly accurate and very reliable: in the various fields observed with the Johnson V filter if a single star with good SNR is taken as a reference and then the others are measured, the overall coincidence with APASS V passband was always within three hundredths of a magnitude regardless of the color index. This is definitely the result of higher quality that I have had so far in determining a star magnitude on-the-fly: the immediate reproducibility of the measurements is very high. In other words, this means that the overall instrumental response of my setup is very similar to that used in the APASS survey.

When you decide to use a star as a reference you must be certain that not only its magnitude has been measured accurately, but also that its brightness is constant over time. The APASS data are not of great help in this case: each star was observed for a few times and any changes are not easily detectable. The good news is that in this case the key factor is not accuracy but precision. You can then use the data from other surveys as NSVS (Wozniak *et al.* 2004), ASAS (Pojmański 2002), and SuperWASP (Butters *et al.* 2010) to assess whether the star has long-term variations. In this work, the comparison stars were chosen in relation to the presence of a good SNR, proximity to the variable star measured, photometric stability in the survey mentioned above, and measurement in the APASS V passband with small uncertainty. As a result all the check stars used in this work showed a photometric stability within two hundredths of a magnitude over the period considered and very good agreement with the APASS measurements.

## 5. Magnitude determination and period calculation

The star's brightness was measured with MAXIMDL v5.23 software (Diffraction

Limited 2012), using the aperture ring method. Since a FWHM of the observing sessions at times arrived at values of 5″ it was decided to choose values providing an adequate signal-to-noise ratio and the certainty of being able to properly contain the whole flux from the star. Usually I used the following apertures: Aperture radius, 11 to 14 pixels; Gap width, 2 to 28 pixels; Annulus thickness, 8 pixels.

Before proceeding further in the analysis, the time of the light curves obtained was heliocentrically corrected (HJD) in order to ensure a perfect compatibility of the data with observations carried out even at a considerable distance in time. The determination of the period was calculated using the software PERIOD04 (Lenz and Breger 2005), using a Discrete Fourier Transform (DFT). The average zero-point (average magnitude of the object) was subtracted from the dataset to prevent the appearance of artifacts centered at a frequency 0.0 of the periodogram. The calculation of the uncertainties was carried out with PERIOD04 using the method described in Breger *et al.* (1999) .

To improve the period determination, NSVS and SWASP photometric data were used when available. However, due to the high scattering which in some cases affects them, the data with high uncertainties were eliminated.  NSVS and SWASP data were also corrected in their zero-point to make them compatible with my V-band standardized data. Having the same zero-point is indeed crucial for correct calculation of the Discrete Fourier Transform operated by PERIOD04.

## 6. Visual inspection of the light curves: finding candidate variable stars

The search for new variable stars in the examined fields followed an approach designed to maximize the probability of success by avoiding long photometric sessions in areas without detectable variable stars. This approach distinguishes between two different ways of working: the discovery-night and the follow-up-night. In the discovery-night the chosen field is observed for the first time as long as possible, taking care to choose a night with the best photometric conditions. At the end of the session the collected data are analyzed with a classic Magnitude-RMS diagram for the presence of variables with noticeable changes. Then each light curve of the stars present in the field up to magnitude 16 V is evaluated visually for other possible variable candidates that are not immediately detectable by the Magnitude-RMS diagram. The usefulness of the proposed method will be explained below in this section.

If the field is interesting a series of follow-up night exposures is planned for the study of candidate variables, otherwise a new field is chosen. It is important to remember that in any case every follow-up-night exposure is checked again with the Magnitude-RMS diagram looking for possible new large variations not detected in the discovery night frames (such as an eclipsing variable that was not varying in the discovery-night), while the visual analysis of the light curves is performed only on the candidate variables detected in the discovery-night, and not for all other stars because of the very long time that this analysis entails.

It is now necessary to clarify what is meant by visual analysis of the light curves, explaining the usefulness of this approach in the discovery of new variables: in Figures 1 and 2 are shown the Magnitude-RMS diagrams of the two considered fields relative to the discovery-night and the position of the variables detected in this survey. By analyzing the graphs, after having obtained all the data presented in this survey, it can be seen at once that all the discovered variables had RMS above average in the discovery-night; the suspected variable NSV 364 (point 8 in Figure 1) that was found to be constant is positioned differently from the others in Figure 1.

It is right to point out that this evidence is very different at the beginning of this survey: the graphs show indeed many other points with high RMS and many of the discovered variables are above average RMS but not so much as to be immediately identifiable. In other words, the graph is more useful to say that the variables are genuine at the end of the survey rather than to discover variables to the limit of the instrument capabilities in a photometric session.

In effect, the construction of a Magnitude-RMS diagram normally occurs with the use of automatic aperture photometry for all the stars in the field: in a field also only moderately crowded, aperture rings that are good for a star are not good for another star due to contamination phenomena.

For example, a star could present within its ring aperture only a portion of a nearby star as well for the ring relative to the sky-background. This is a great source of 5–6% flux scatter. The result is that in practice many non-variable stars have a scatter higher than other real variable stars of the same magnitude. At the end, they become indistinguishable in the diagram considered.

Visual analysis of the light curve allows instead to evaluate not only a significant change in the typical scatter but also the presence of shapes and meaningful patterns in the light curve. For every star the best aperture ring, gap ring, and sky-background ring are determined, even for the comparison stars. This type of evaluation is impossible with a diagram Magnitude-RMS: there are too many things to consider in order to be included in an automatic algorithm.

Figure 3 shows a graph that allows to understand the value of a visual assessment of one-night light curve of the variable star 2MASS J00584964 +5909260, probably the star with the most critical amplitude variation-SNR of this survey. Note that in the latter part of the night faint clouds appeared at high altitude. In Figure 3 the minimal variation coupled with a pulsation-shape of the light curve is enough to consider the star as a potential variable candidate. Of course it would also be folly to think that this is enough to consider the star as a real variable: in these cases the path of analysis must be very thorough, since the variation could also be only noise.

In order to distinguish among variables, suspected variables, and non-variables we proceeded as follows. To be considered as a variable the star must have shown similar variations in every night and these variations must be consistent in the final phase diagram. Most importantly the variability must be confirmed in another dataset. In fact, all the stars discovered in this survey (except GSC 04051-02483, 2MASS J02472793 +6149024, GSC 04047-00381, 2MASS J02443720+6143091, and GSC 04047-01418 due to the lack of available data) were cross-checked against the SWASP and NSVS datasets in order to confirm the variability and improve the determination of the period by combining the available data with the data obtained in this survey. The phase diagrams presented at the end of this work show the superposition of the various datasets and their coincidence.

If a phase diagram highlights a variation comparable to the uncertainty and there are no other datasets available to perform a cross-check, the high cadence and the large number of measurements made in this survey allow you to make a binning to improve the SNR. If also in this case the variation is uncertain the star is only considered a potential variable. No star presented in this work is still in this condition.

## 7. New variable stars and variability checking of NSV 364

In this survey we discovered eighteen new variable stars and checked the suspected variable NSV 364 for variability. The population of the new variables is as follows:

- 8 pulsating (7 δ Sct and 1γ Dor)
- 6 eclipsing (1 β Lyr, 2 W UMa, and 3 β Per)
- 3 rotating (3 rotating ellipsoidal)
- 1 eruptive (1 Be with LERI variations).

The coordinates of all new variable stars discovered in this survey are reported as they appear in the UCAC4 catalogue (Zacharias *et al.* 2012) and differ from the detected positions for a value never greater than 0.5".

### 7.1. GSC 03680-00423

Position (UCAC4): R.A. (J2000) = $01^h 03^m 51.302^s$, Dec. (J2000) = +59° 18' 57.83"

Cross Identification: 1SWASP J010351.26+591858.5; 2MASS J01035130+5918577; UCAC3 299-021274; UCAC4 747-009468; USNO-B1.0 1493-0034696

Variability Type: β Lyr

Magnitude: Max. 13.10 V; Min. 13.37 V (Secondary max. 13.13 V; Secondary min. 13.28 V)

Period: 0.509944(1) d

Epoch: 2456223.2844(22) HJD

Comparison Star: UCAC4 747-009434 (APASS 12.775 V)

Check Star: UCAC4 747-009352

Finding chart, phase plot, and Fourier spectrum are shown in Figures 4, 5, and 6.

7.2. GSC 03680-00667

Position (UCAC4): R.A. (J2000) = $01^h 03^m 20.400^s$, Dec. (J2000) = +59° 13' 45.46"

Cross Identification: 1SWASP J010320.40+591345.6;
2MASS J01032039+5913453; UCAC4 747-009355;
USNO-B1.0 1492-0033355

Variability Type:  δ Sct

Magnitude: Max. 13.47 V; Min. 13.52 V

Period: 0.1584384(6) d

Epoch: 2456230.399(2) HJD

Comparison Star: UCAC4 747-009434 (APASS 12.775 V)

Check Star: UCAC4 747-009352

Notes: Despite the wide time span of the data used for the period determination (SWASP + Furgoni) the Fourier spectrum shows two significant peaks at a frequency of 6.31160976 c/d and 7.31438512 c/d that differ only by a width of 0.0007 mag. One of the two peaks is probably the alias of the other but the two very similar amplitudes do not allow determination with absolute certainty of the correct one. For this reason the period may also be P=0.1367168(4).

Finding chart, phase plot, and Fourier spectrum are shown in Figures 7, 8, and 9.

7.3. 2MASS J01020513+5912394

Position (UCAC4): R.A. (J2000) = $01^h 02^m 05.134^s$, Dec. (J2000) = +59° 12' 39.36"

Cross Identification: 1SWASP J010205.11+591239.5;
UCAC4 747-009100; USNO-B1.0 1492-0032522

Variability Type: Rotating ellipsoidal

Magnitude: Max. 13.91 V; Min. 14.04 V

Period: 0.873754(2) d

Epoch: 2456223.4065(21) HJD

Comparison Star: UCAC4 746-009480 (APASS 11.045 V)

Check Star: UCAC4 746-009545

Finding chart, phase plot, and Fourier spectrum are shown in Figures 10, 11, and 12.

7.4. GSC 03680-01488

Position (UCAC4): R.A. (J2000) = 00$^h$ 58$^m$ 33.793$^s$, Dec. (J2000) = +58° 57' 17.87"

Cross Identification: 1SWASP J005833.79+585717.8;
2MASS J00583378+5857178; USNO-B1.0 1489-0030192

Variability Type: δ Sct

Magnitude: Max. 11.155 V; Min. 11.180 V

Main Period: 0.046419(1) d

Secondary Period: 0.0938359(2) d

Epoch Main Period: 2456223.3449(5) HJD

Epoch Secondary Period: 2456223.3454(12) HJD

Ensemble Comparison Stars: UCAC4 745-008614 (APASS 11.006 V);
UCAC4 745-008750 (APASS 11.569 V); UCAC4 745-008771
(APASS 11.663 V)

Check Star: UCAC4 745-008825

Notes: The light curve shows an evident modulation in the different nights of observation. The Fourier spectrum shows the existence of a possible other active frequency that is very close to half the main frequency (p = 0.0938359 d).

Finding chart, phase plots, and Fourier spectrum are shown in Figures 13, 14, 15, and 16.

7.5. GSC 03680-01320

Position (UCAC4): R.A. (J2000) = 01$^h$ 01$^m$ 39.227$^s$, Dec. (J2000) = +58° 55' 44.66"

Cross Identification: 1SWASP J010139.22+585544.8;
2MASS J01013922+5855447; UCAC4 745-008924;
USNO-B1.0 1489-0031868

Variability Type:  δ Sct

Magnitude: Max. 13.035 V; Min. 13.060 V

Main Period: 0.0631941(1) d

Secondary Period: 0.0454240(1) d

Epoch Main Period: 2456223.2334(11) HJD

Epoch Secondary Period: 2456223.2798(9) HJD

Ensemble Comparison Stars: UCAC4 745-009004 (APASS 11.006 V);
UCAC4 745-009144 (APASS 11.722 V)

Check Star: UCAC4 745-008982

Notes: The continuous modulation of the light curve suggests the existence of different active frequencies. The secondary frequency corresponds to a P =

0.0454240(1) with $HJD_{max}$ = 2456223.2798(9) and an amplitude slightly lower than that of the main period. The analysis of the Fourier spectrum shows in any case that additional frequencies are probably active.

Finding chart, phase plots and Fourier spectrum are shown in Figures 17, 18, 19, and 20.

### 7.6. 2MASS J00584964+5909260

Position (UCAC4): R.A. (J2000) = $00^h 58^m 49.657^s$, Dec. (J2000) = +59° 09' 26.09"

Cross Identification: 1SWASP J005849.66+590926.4; UCAC4 746-008985; USNO-B1.0 1491-0031750

Variability Type: δ Sct

Magnitude: Max. 14.11; Min. 14.13 V

Period: 0.193186(2) d

Epoch: 2456223.2446(38) HJD

Ensemble Comparison Stars: UCAC4 747-008816 (APASS 11.599 V); UCAC4 746-009298 (APASS 11.646 V)

Check Star: UCAC4 746-009381

Finding chart, phase plot, and Fourier spectrum are shown in Figures 21, 22, and 23.

### 7.7. ALS 6430

Position (UCAC4): R.A. (J2000) = $01^h 00^m 17.527^s$, Dec. (J2000) = +59° 08' 13.34"

Cross Identification: 1SWASP J010017.52+590813.2; 2MASS J01001753+5908133; GSC 03680-01411; LS I +58 20

Variability Type: Rotating ellipsoidal

Magnitude: Max. 11.38 V; Min. 11.41 V

Period: 1.6814956(88) d

Epoch: 2456223.4758(58) HJD

Ensemble Comparison Stars: UCAC4 747-008816 (APASS 11.599 V); UCAC4 746-009298 (APASS 11.646 V)

Check Star: UCAC4 746-009381

Notes: The APASS B-V for this star is 0.308, 2MASS J-K 0.163, redder than expected for the spectral type OB taken from the LS catalog. SPB or ACV type are possible with half the period.

Finding chart, phase plot, and Fourier spectrum are shown in Figures 24, 25, and 26.

7.8. NSV 364

Position (UCAC4): R.A. (J2000)=01$^h$01$^m$23.856$^s$, Dec. (J2000)=+59° 04'32.88"

Cross Identification: 1SWASP J010123.84+590433.1;
2MASS J01012386+5904327; SON 10457; UCAC4 746-009503;
USNO-B1.0 1490-0032435

Variability Type: Constant; Non-Variable

Magnitude: 14.49 V

Comparison Star: UCAC4 746-009480 (APASS 11.045 V)

Check Star: UCAC4 746-009545

Notes: The star is constant both in the Furgoni dataset and the SWASP dataset. The Fourier Power Spectrum of SWASP dataset shows an equal distribution of peaks spaced by 0.5 c/d. No relevant frequencies present. The suspected variability was described in Richter (1969).

Finding chart, light curve, and Fourier spectrum are shown in Figures 27, 28, and 29.

7.9. GSC 04051-02533

Position (UCAC4): R.A. (J2000)=02$^h$47$^m$42.288$^s$, Dec. (J2000)=+61° 58'29.35"

Cross Identification: 1SWASP J024742.26+615829.2;
2MASS J02474229+6158293; UCAC4 760-022284

Variability Type:  β Per

Magnitude: Max. 12.37 V; Min. 12.72 V

Period: 1.57709(1) d

Epoch: 2456265.458(2) HJD

Comparison Star: UCAC4 760-022374 (APASS 12.081 V)

Check Star: UCAC4 761-021189

Notes: The binary system is probably included in the very small DSH J0247.7+6158=Teutsch 162 cluster of stars, involved in the HII-region Sh 2-193. The secondary minimum is probably as deep as the primary one.

Finding chart and phase plot are shown in Figures 30 and 31.

7.10. GSC 04051-02027

Position (UCAC4): R.A. (J2000)=02$^h$46$^m$10.965$^s$, Dec. (J2000)=+61° 57'55.69"

Cross Identification: 1SWASP J024611.00+615755.8;
2MASS J02461097+6157557; UCAC4 760-022038

Variability Type: β Per

Magnitude: Max. 12.67 V; Min. 12.85 V (Secondary min. 12.83 V)

Period: 1.568639(2) d

Epoch: 2456329.1377(36) HJD

Ensemble Comparison Stars: UCAC4 760-022374 (APASS 12.081 V); UCAC4 761-021049 (APASS 12.033 V).

Check Star: UCAC4 761-021189

Finding chart, phase plot, and Fourier spectrum are shown in Figures 32, 33, and 34.

7.11. GSC 04051-01669

Position (UCAC4): R.A. (J2000)=02$^h$48$^m$19.071$^s$, Dec. (J2000)=+61°57'03.19"

Cross Identification: 1SWASP J024819.07+615703.2; 2MASS J02481907+6157031; UCAC4 760-022382

Variability Type: Rotating ellipsoidal

Magnitude: Max. 13.66 V; Min. 13.82 V

Period: 1.695249(6) d

Epoch: 2456266.2219(68) HJD

Ensemble Comparison Stars: UCAC4 760-022374 (APASS 12.081 V); UCAC4 761-021049 (APASS 12.033 V).

Check Star: UCAC4 761-021189

Finding chart, phase plot, and Fourier spectrum are shown in Figures 35, 36, and 37.

7.12. GSC 04051-01789

Position (UCAC4): R.A. (J2000)=02$^h$47$^m$55.761$^s$, Dec. (J2000)=+62°09'06.86"

Cross Identification: 1SWASP J024755.74+620906.9; 2MASS J02475576+6209069; UCAC4 761-021226

Variability Type: δ Sct

Magnitude: Max. 12.25 V; Min. 12.29 V

Period: 0.13367394(11) d

Epoch: 2456268.37961(78) HJD

Ensemble Comparison Stars: UCAC4 760-022374 (APASS 12.081 V); UCAC4 761-021049 (APASS 12.033 V).

Check Star: UCAC4 761-021189

Finding chart, phase plot, and Fourier spectrum are shown in Figures 38, 39, and 40.

7.13. GSC 04051-02483

Position (UCAC4): R.A. (J2000) = 02$^h$46$^m$06.407$^s$, Dec. (J2000) = +61° 54'23.93"

Cross Identification: 2MASS J02460641+6154239; EM* CDS 304;
LS I +61 312; NSVS 1883898; UCAC4 760-022024

Variability Type: Be + LERI

Magnitude: Max. 11.52 V; Min. 11.63 V

Period: 0.25573(3) d

Epoch: 2456267.147(3) HJD

Ensemble Comparison Stars: UCAC4 760-022374 (APASS 12.081 V);
UCAC4 761-021049 (APASS 12.033 V).

Check Star: UCAC4 761-021189

Notes: H-$\alpha$ emission-line star (the spectral type is B5IIIe (Skiff 2009–2013)).
Fading event detected on 6 December 2013. The small amplitude irregular
variability (Be) is coupled with quasi-periodic variations of the LERI-type.
The period could be 0.51146 d with a double-waved light curve. Epoch of
maximum given.

Finding chart, light curves, phase plot, and Fourier spectrum are shown in
Figures 41, 42, 43, 44, and 45.

7.14. 2MASS J02472793+6149024

Position (UCAC4): R.A. (J2000) = 02$^h$47$^m$27.936$^s$, Dec. (J2000) = +61° 49'02.46"

Cross Identification: UCAC4 760-022245; USNO-B1.0 1518-0082596

Variability Type: δ Sct

Magnitude: Max. 14.85 V; Min. 14.94 V

Main Period: 0.132260(8) d

Secondary Period: 0.092002(4)

Epoch Main Period: 2456268.3360(15) HJD

Epoch Secondary Period: 2456268.3237(10) HJD

Ensemble Comparison Stars: UCAC4 760-022283 (APASS 14.465 V);
UCAC4 760-022267 (APASS 14.691 V).

Check Star: UCAC4 759-022016

Notes: B–V = 0.95 (APASS) probably reddened. 2MASS J02472833+6149073
(V = 17.4) lies 5.6" away and has not been included in the photometry.

Finding chart, phase plots, and Fourier spectrum are shown in Figures 46, 47,
48, and 49.

7.15. GSC 04047-01118

Position (UCAC4): R.A. (J2000)=02$^h$48$^m$31.004$^s$, Dec. (J2000)=+61°40'25.48"

Cross Identification: 2MASS J02483099+6140255; IC 1848 +61 41;
NSVS 1885359; TYC 4047-1118-1; UCAC4 759-022220

Variability Type: γ Dor

Magnitude: Max. 11.84 V; Min. 11.88 V

Period: 0.4065110(8) d

Epoch: 2456265.4098(29) HJD

Ensemble Comparison Stars: UCAC4 759-021947 (APASS 12.233 V);
UCAC4 759-022035 (APASS 12.530 V).

Check Star: UCAC4 759-021930

Notes: B–V 0.564 (APASS). The star is in the open cluster and nebula
IC 1848.

Finding chart, phase plot, and Fourier spectrum are shown in Figures 50, 51, and 52.

7.16. GSC 04047-00558

Position (UCAC4): R.A. (J2000)=02$^h$46$^m$26.433$^s$, Dec. (J2000)=+61°36'54.44"

Cross Identification: 2MASS J02462643+6136545; UCAC4 759-021892

Variability Type: β Per

Magnitude: Max. 14.25 V; Min. 14.62 V (Secondary min. 14.52 V)

Period: 1.96466(3) d

Epoch: Ensemble Comparison Stars: UCAC4 759-021947 (APASS 12.233 V);
UCAC4 759-022035 (APASS 12.530 V).

Check Star: UCAC4 759-021930

Finding chart and phase plot are shown in Figures 53 and 54.

7.17. GSC 04047-00381

Position (UCAC4): R.A. (J2000)=02$^h$44$^m$18.468$^s$, Dec. (J2000)=+61°45'04.53"

Cross Identification: 2MASS J02441846+6145045; UCAC4 759-021534

Variability Type: W UMa

Magnitude: Max. 14.76 V; Min. 15.10 V

Period: 0.348687(3) d

Epoch: 2456267.3401(5) HJD

Ensemble Comparison Stars: UCAC4 760-021845 (APASS 12.487 V);
UCAC4 760-021940 (APASS 12.025 V).

Check Star: UCAC4 759-021771

Finding chart, phase plot, and Fourier spectrum are shown in Figures 55, 56, and 57.

### 7.18. 2MASS J02443720+6143091

Position (UCAC4): R.A. (J2000) = $02^h 44^m 37.209s$, Dec. (J2000) = +61°43'09.07"

Cross Identification: UCAC4 759-021583

Variability Type: W UMa

Magnitude: Max. 15.12 V; Min. 15.49 V

Period: 0.353443(5) d

Epoch: 2456268.3037(7) HJD

Ensemble Comparison Stars: UCAC4 760-021845 (APASS 12.487 V); UCAC4 760-021940 (APASS 12.025 V).

Check Star: UCAC4 759-021771

Finding chart, phase plot, and Fourier spectrum are shown in Figures 58, 59, and 60.

### 7.19. GSC 04047-01418

Position (UCAC4): R.A. (J2000) = $02^h 45^m 11.691s$, Dec. (J2000) = +61°43'53.78"

Cross Identification: 2MASS J02451169+6143538; UCAC4 759-021671

Variability Type: δ Sct

Magnitude: Max. 14.13 V; Min. 14.16 V

Period: 0.17623(2) d

Epoch: 2456265.316(3) HJD

Ensemble Comparison Stars: UCAC4 760-021845 (APASS 12.487 V); UCAC4 760-021940 (APASS 12.025 V).

Check Star: UCAC4 759-021771

Finding chart, phase plot, and Fourier spectrum are shown in Figures 61, 62, and 63.

## 8. Acknowledgements

I wish to thank Sebastian Otero, member of the VSX team and AAVSO external consultant, for his helpful comments on the new variable stars discovered. This work has made use of the VizieR catalogue access tool, CDS, Strasbourg, France, and the International Variable Star Index (VSX) operated by the AAVSO. This work has made use of the ASAS3 Public Catalogue (Pojmański *et al.* 2013), NSVS data obtained from the Sky Database for Objects in Time-Domain operated by the Los Alamos National Laboratory, and data obtained from the SuperWASP Public Archive operated by the WASP consortium, which consists of representatives from the Queen's University Belfast, the University of Cambridge (Wide Field Astronomy Unit), Instituto de Astrofisica de Canarias, the Isaac Newton Group of Telescopes (La Palma), the University of Keele, the University of Leicester, the Open University, the University of St. Andrews, and the South African Astronomical Observatory.

This work has made use of The Fourth U.S. Naval Observatory CCD Astrograph Catalog (UCAC4).

Table 1. Dates and times of observations conducted in this study.

| | | | |
|---|---|---|---|
| *Field R.A. 01ʰ 01ᵐ 12ˢ Dec. +59° 05′ 11″* | | | |
| *Date (dd-mm-yyyy)* | *UTC Start (hh:mm:ss)* | *UTC End (hh:mm:ss)* | *Useful Number of Exposures* |
| 22-10-2012 | 18:13:29 | 22:35:31 | 120 |
| 23-10-2012 | 17:31:24 | 21:57:10 | 120 |
| 24-10-2012 | 17:21:23 | 21:53:21 | 120 |
| 29-10-2012 | 17:14:07 | 22:08:48 | 135 |
| 01-11-2012 | 19:28:32 | 22:50:03 | 67 |
| 05-11-2012 | 17:09:34 | 22:13:12 | 140 |
| 06-11-2012 | 17:38:27 | 22:11:26 | 126 |
| *Field R.A. 02ʰ 46ᵐ 09ˢ Dec. +61° 53′ 12″* | | | |
| *Date (dd-mm-yyyy)* | *UTC Start (hh:mm:ss)* | *UTC End (hh:mm:ss)* | *Useful Number of Exposures* |
| 03-12-2012 | 16:56:13 | 22:25:51 | 151 |
| 05-12-2012 | 17:10:10 | 22:25:06 | 140 |
| 06-12-2012 | 17:23:53 | 22:25:12 | 135 |
| 10-12-2012 | 18:58:35 | 23:33:11 | 101 |
| 22-01-2013 | 18:04:15 | 23:20:00 | 144 |
| 26-01-2013 | 17:36:09 | 20:00:16 | 66 |
| 04-02-2013 | 18:08:29 | 22:50:20 | 129 |
| 10-02-2013 | 17:37:00 | 18:31:30 | 25 |
| 14-02-2013 | 18:38:56 | 21:20:30 | 74 |
| 18-02-2013 | 17:59:39 | 23:27:56 | 151 |
| 19-02-2013 | 17:50:28 | 21:48:21 | 103 |
| 26-02-2013 | 18:02:47 | 20:46:25 | 75 |
| 28-02-2013 | 18:00:17 | 21:04:39 | 81 |
| 04-03-2013 | 18:05:32 | 22:54:27 | 126 |

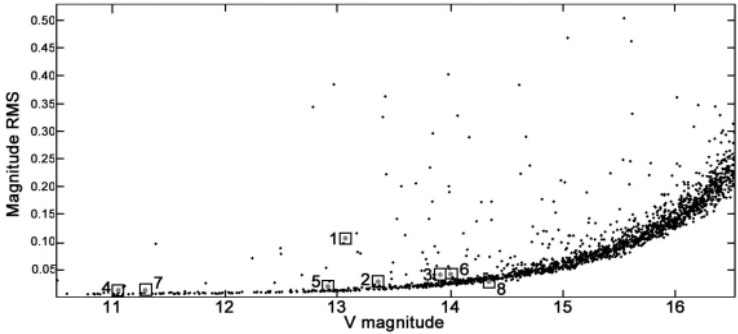

Figure 1. Magnitude-RMS diagram for the discovery-night (22-10-2012) of the field R.A. 01ʰ 01ᵐ 12ˢ, Dec. +59° 05' 11". The squares are the variable stars detected in this survey except for No. 8 which is NSV 364 and was found to be non-variable.

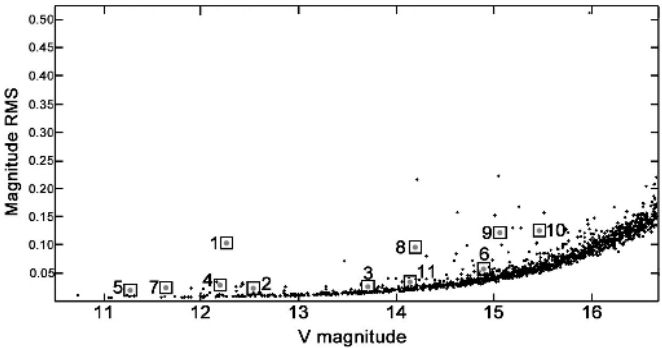

Figure 2. Magnitude-RMS diagram for the discovery-night (03-12-2012) of the field R.A. 02ʰ 46ᵐ 09ˢ, Dec. +61° 53'1 2". The squares are the variable stars detected in this survey.

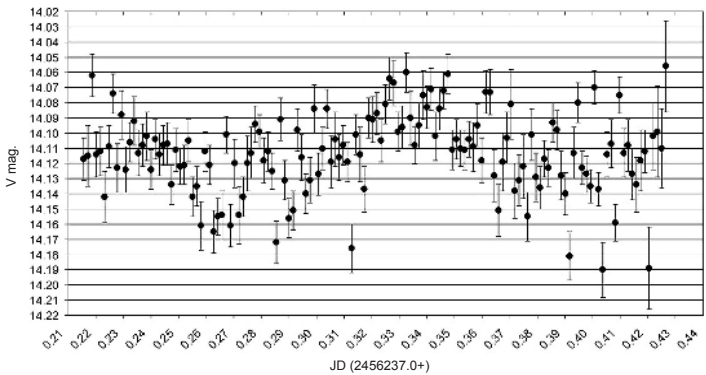

Figure 3. One night light curve of 2MASS J00584964+5909260.

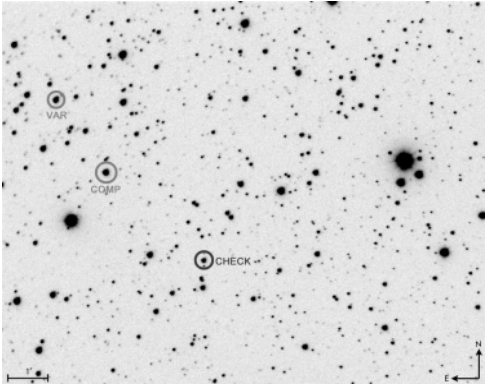

Figure 4. Finding chart of GSC 03680-00423.

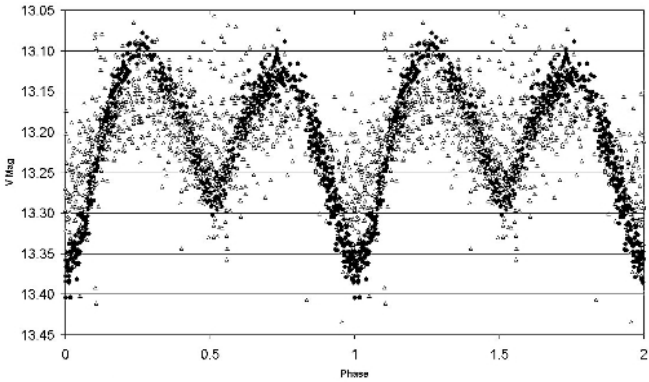

Figure 5. Phase plot of GSC 03680-00423. Filled circles denote Furgoni data; open triangles denote SWASP data (different zero-point corrections applied).

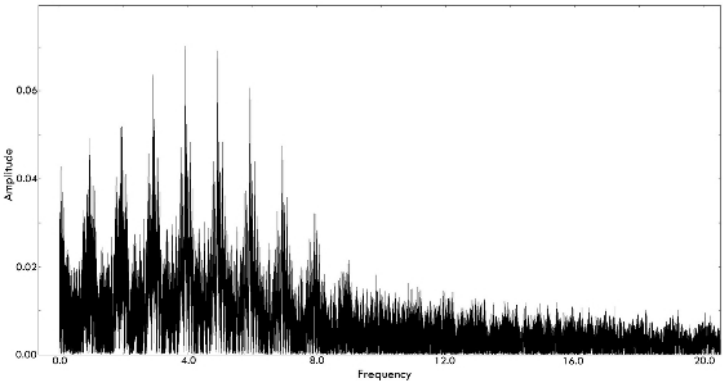

Figure 6. Fourier spectrum of GSC 03680-00423.

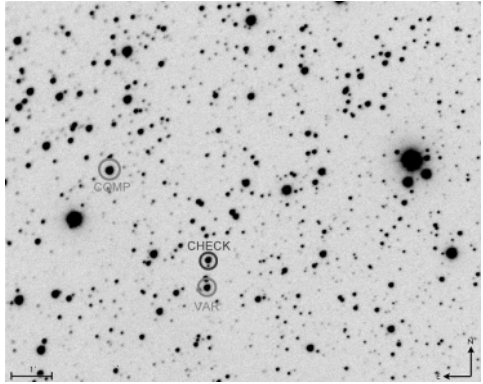

Figure 7. Finding chart of GSC 03680-00667.

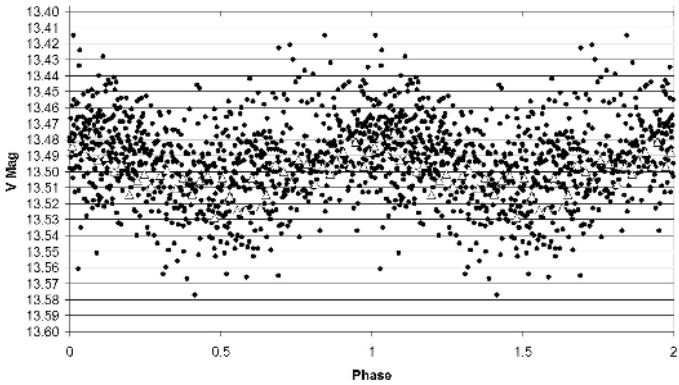

Figure 8. Phase plot of GSC 03680-00667. Filled circles denote Furgoni dataset; open triangles denote SWASP dataset with error less than 0.1 mag. +0.35 mag. offset applied.

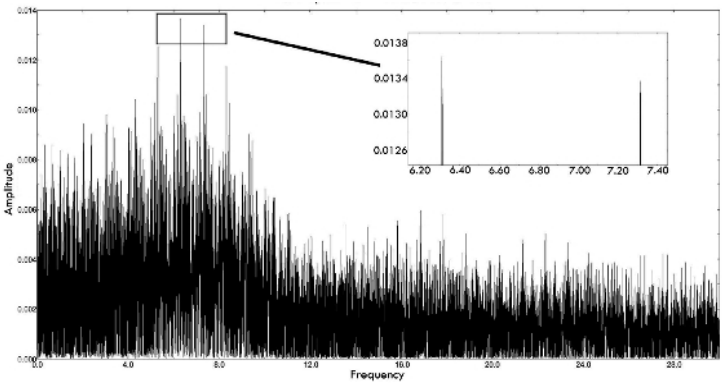

Figure 9. Fourier spectrum of GSC 03680-00667.

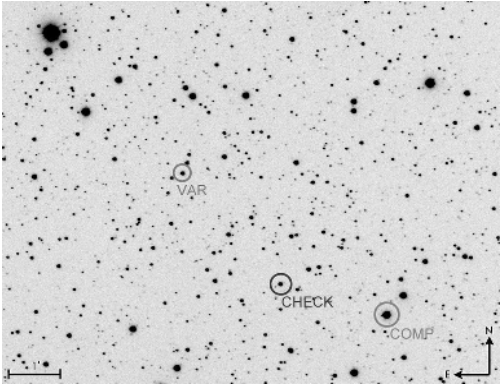

Figure 10. Finding chart of 2MASS J01020513+5912394.

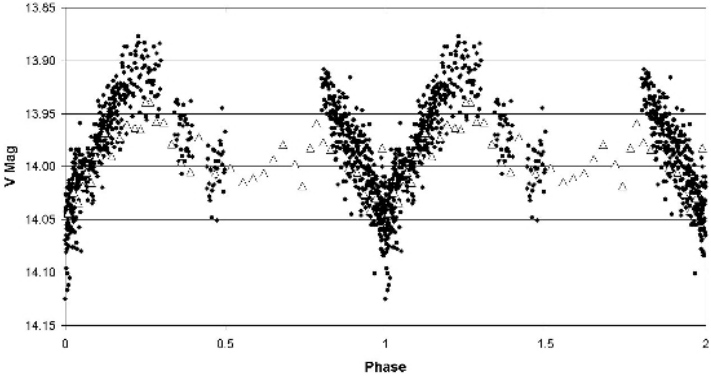

Figure 11. Phase plot of 2MASS J01020513+5912394. Filled circles denote Furgoni dataset; open triangles denote SWASP data with error less than 0.03 mag. (+0.57 mag. offset applied, 15 points binning).

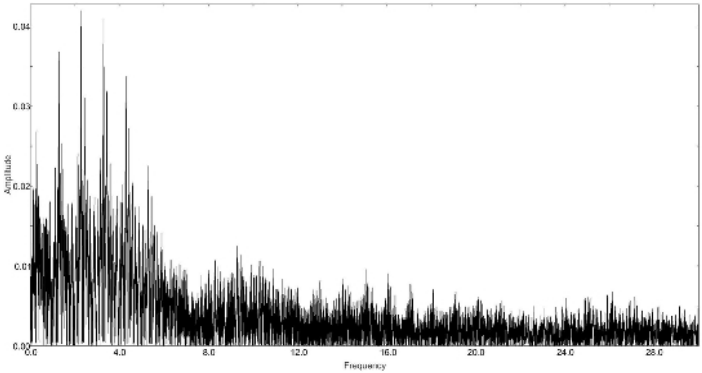

Figure 12. Fourier spectrum of 2MASS J01020513+5912394.

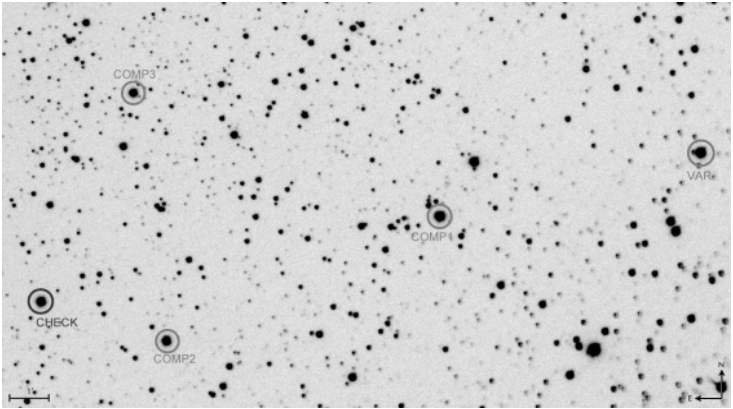

Figure 13. Finding chart of GSC 03680-01488.

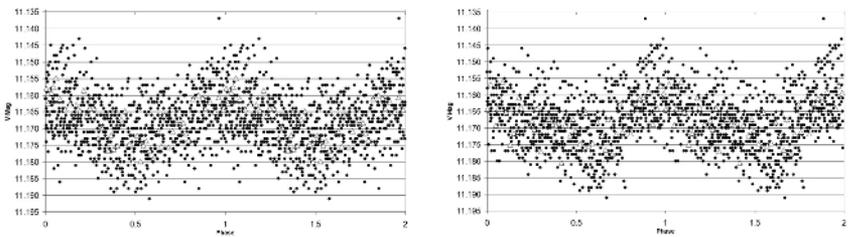

Figures 14 and 15. Main (left) and secondary (right) period phase plots of GSC 03680-01488. Filled circles denote Furgoni dataset; open triangles denote SWASP dataset with error less than or equal to 0.01 mag. Binning 15 and – 0.125 mag. offset applied.

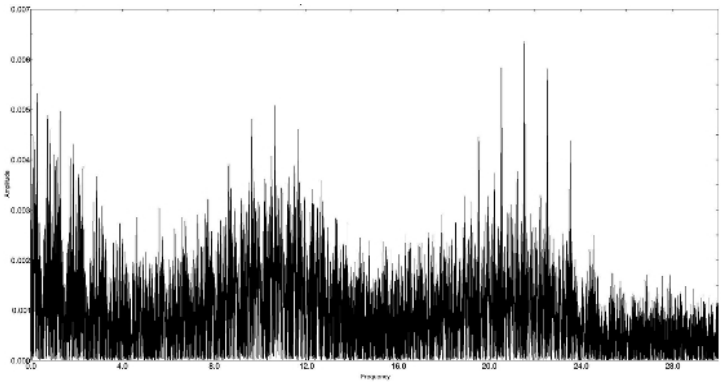

Figure 16. Fourier spectrum of GSC 03680-01488.

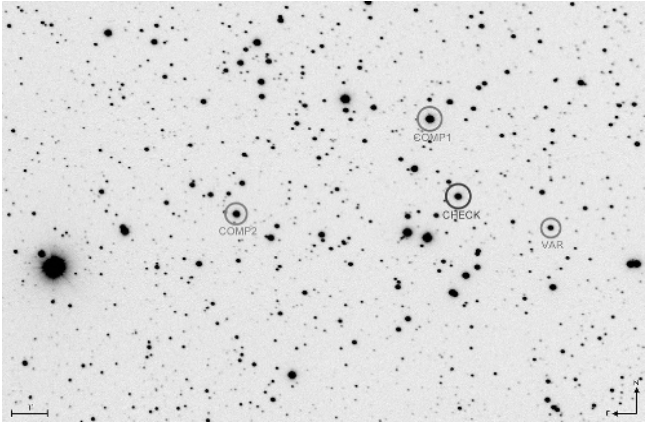

Figure 17. Finding chart of GSC 03680-01320.

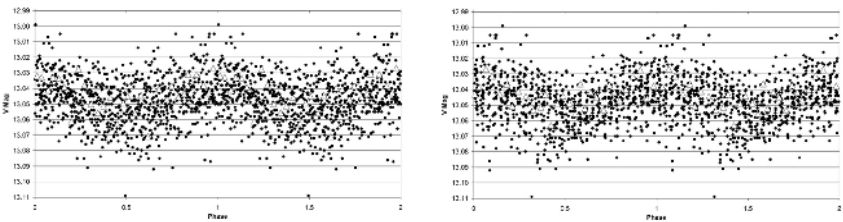

Figures 18 and 19. Main (left) and secondary (right) period phase plots of GSC 03680-01320. Filled circles denote Furgoni dataset; open triangles denote SWASP dataset with error less than 0.04 mag. and –0.35 mag. offset applied (Binning 30).

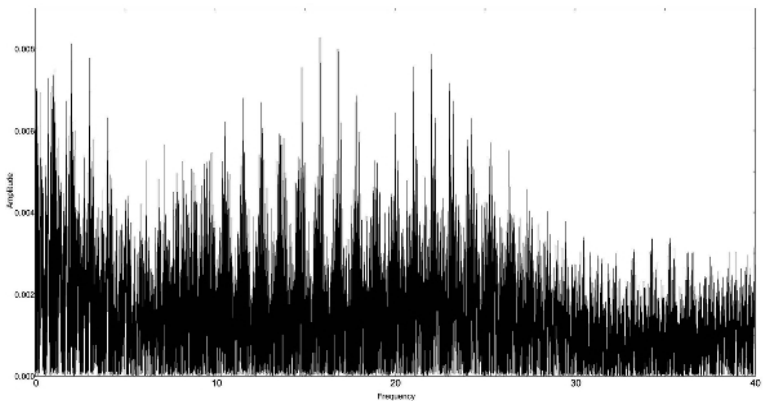

Figure 20. Fourier spectrum of GSC 03680-01320.

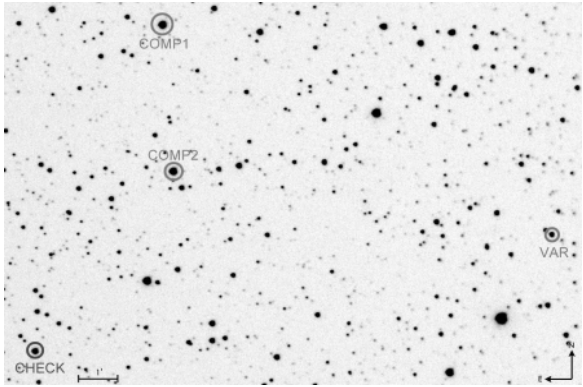

Figure 21. Finding chart of 2MASS J00584964+5909260.

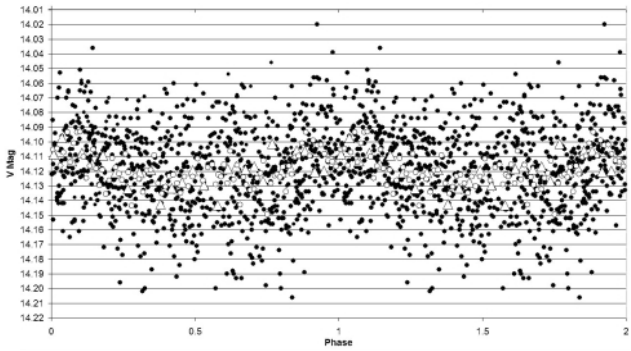

Figure 22. Phase plot of 2MASS J00584964+5909260. Filled circles denote Furgoni dataset; open triangles denote SWASP dataset with error less than 0.06 mag. Binning 30 and −0.35 mag. offset applied. Open circles denote Furgoni dataset, Binning 10.

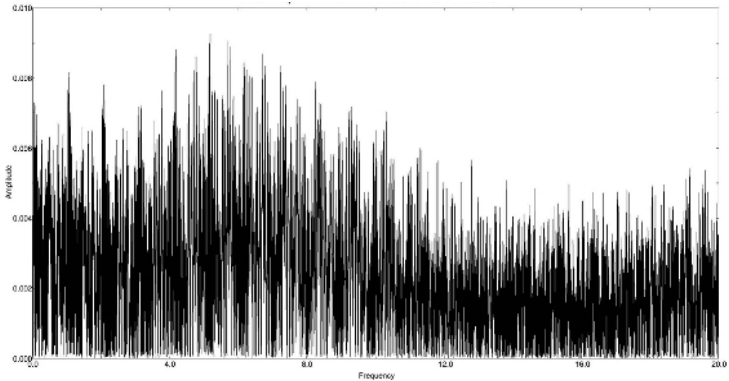

Figure 23. Fourier spectrum of 2MASS J00584964+5909260.

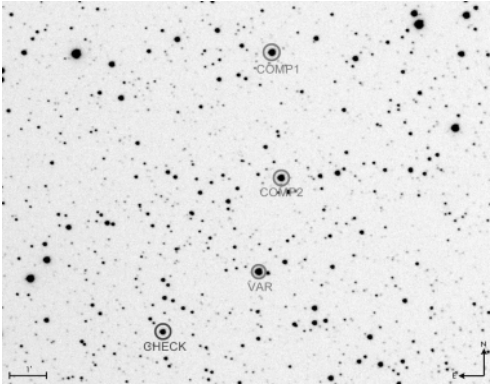

Figure 24. Finding chart of ALS 6430.

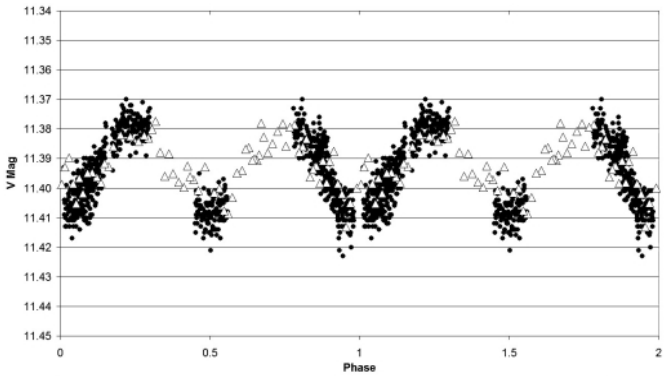

Figure 25. Phase plot of ALS 6430. Filled circles denote Furgoni dataset; open triangles denote SWASP dataset with errors less than 0.03 mag., binning 20, –0.100 mag. offset applied.

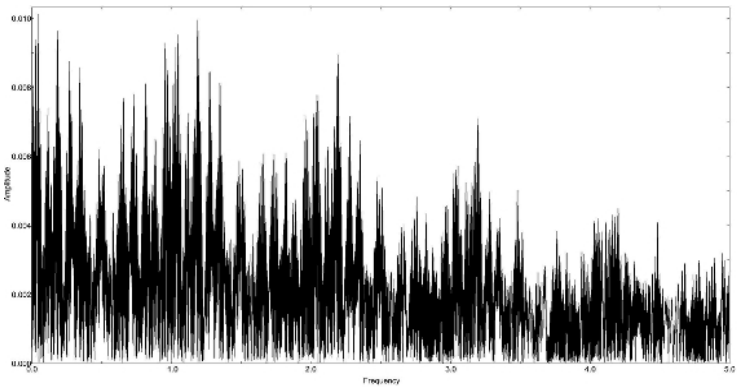

Figure 26. Fourier spectrum of ALS 6430.

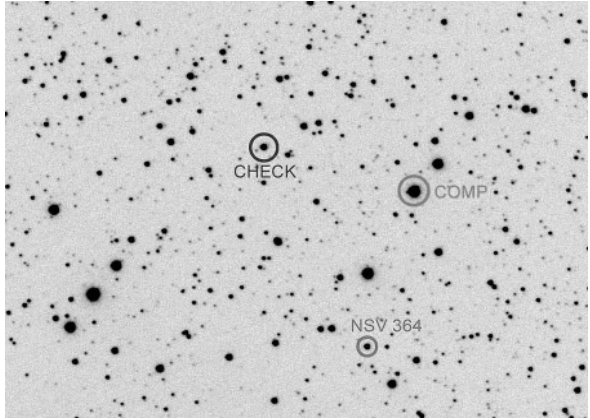

Figure 27. Finding chart of NSV 364.

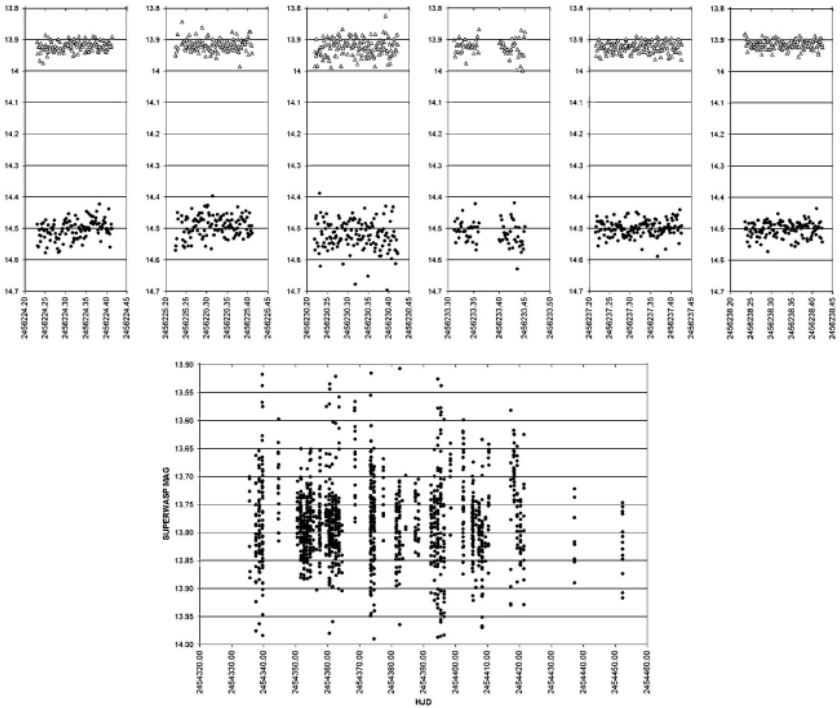

Figure 28. Light curves of NSV 364. In the Furgoni dataset (upper plot), open triangles denote the check star, and filled circles denote NSV 364; x-axis is JD, y-axis is V mag. Light curve of the SuperWasp dataset is shown in the lower plot; x-axis is HJD, y-axis is SWASP magnitude.

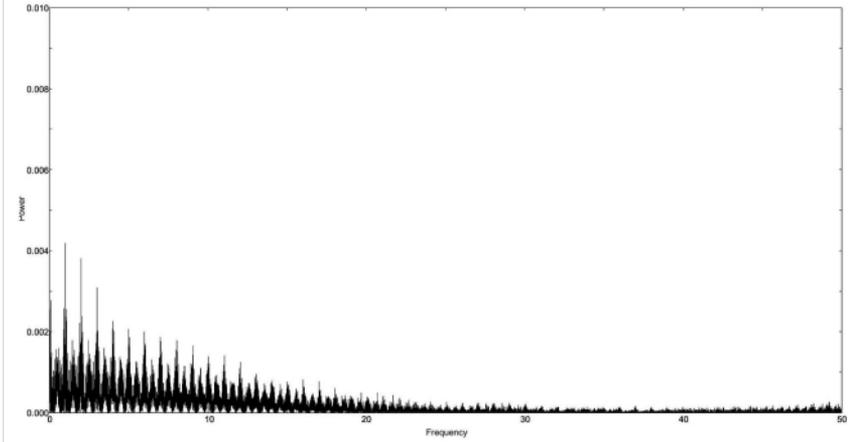

Figure 29. Fourier spectrum of NSV 364.

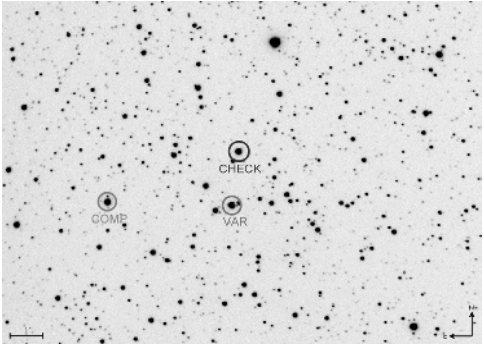

Figure 30. Finding chart of GSC 04051-02533.

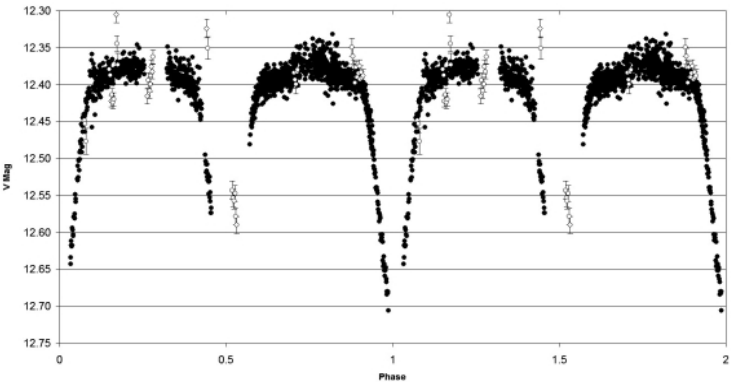

Figure 31. Phase plot of GSC 04051-02533. Filled circles denote Furgoni dataset; open circles denote SWASP dataset (−0.40 mag. offset).

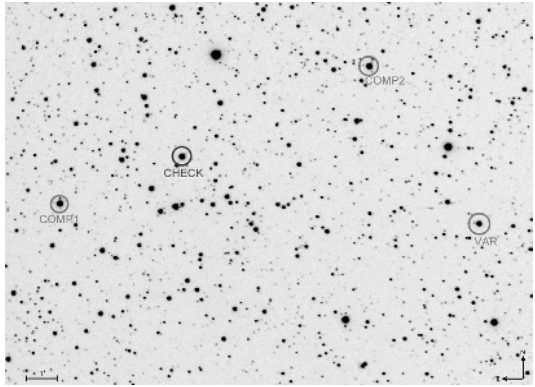

Figure 32. Finding chart of GSC 04051-02027.

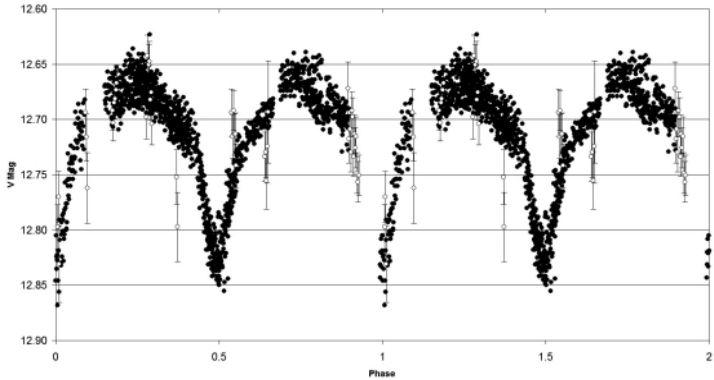

Figure 33. Phase plot of GSC 04051-02027. Filled circles denote Furgoni dataset; open circles denote SWASP dataset (–0.21 mag. offset).

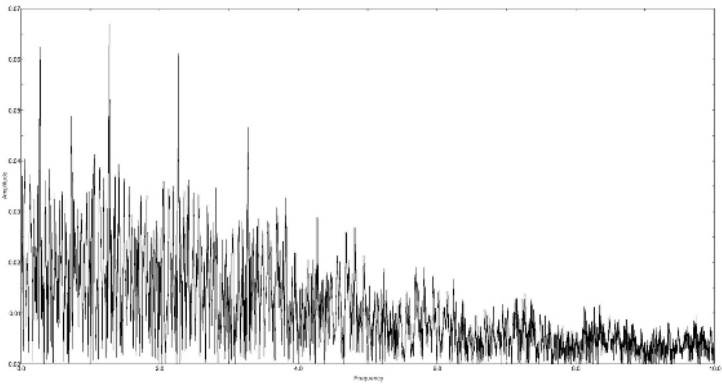

Figure 34. Fourier spectrum of GSC 04051-02027.

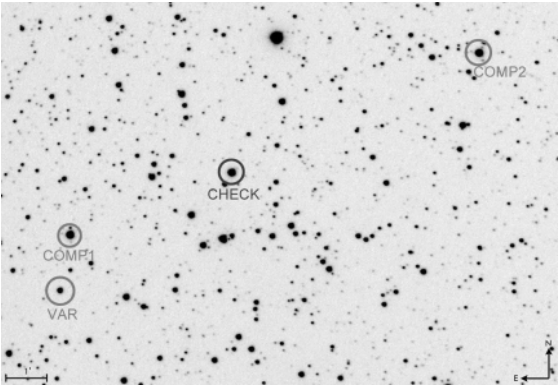

Figure 35. Finding chart of GSC 04051-01669.

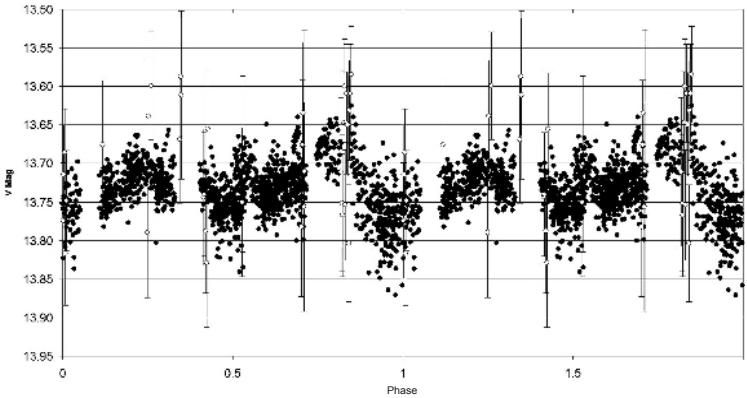

Figure 36. Phase plot of GSC 04051-01669. Filled circles denote Furgoni dataset; open circles denote SWASP dataset (–0.78 mag. offset).

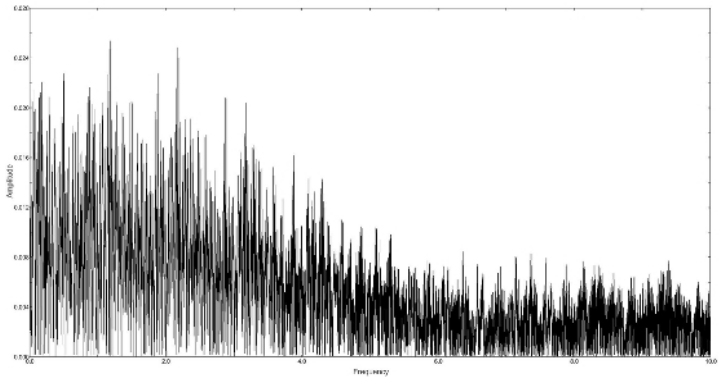

Figure 37. Fourier spectrum of GSC 04051-01669.

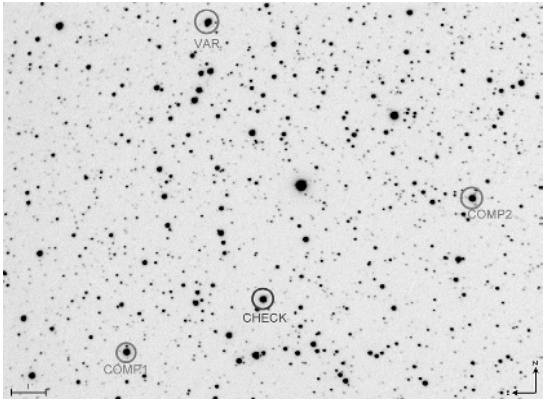

Figure 38. Finding chart of GSC 04051-01789.

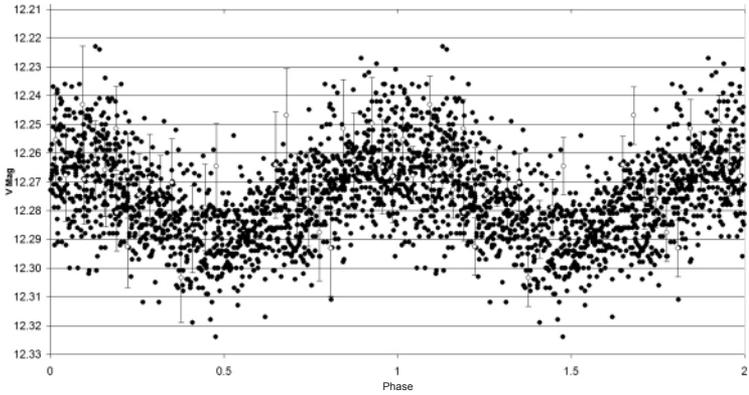

Figure 39. Phase plot of GSC 04051-01789. Filled circles denote Furgoni dataset; open circles denote SWASP dataset (–0.05 mag. offset).

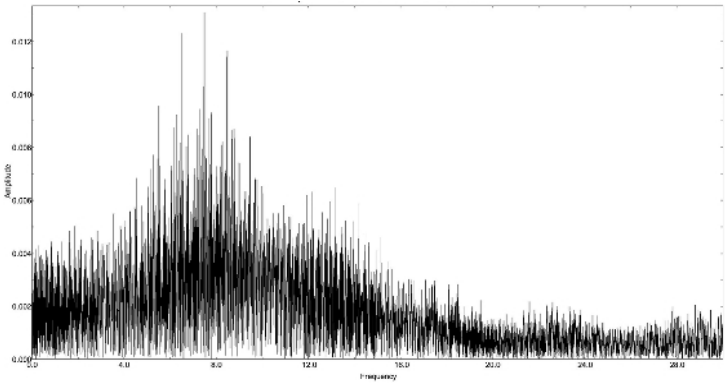

Figure 40. Fourier spectrum of GSC 04051-01789.

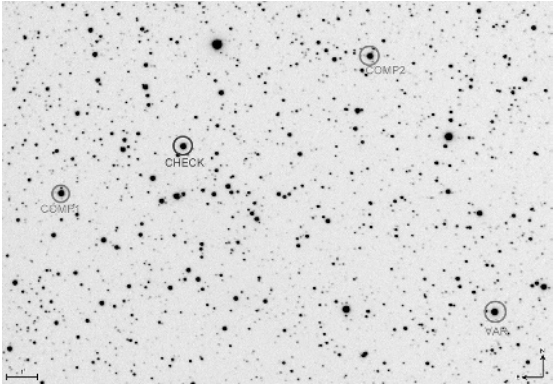

Figure 41. Finding chart of GSC 04051-02483.

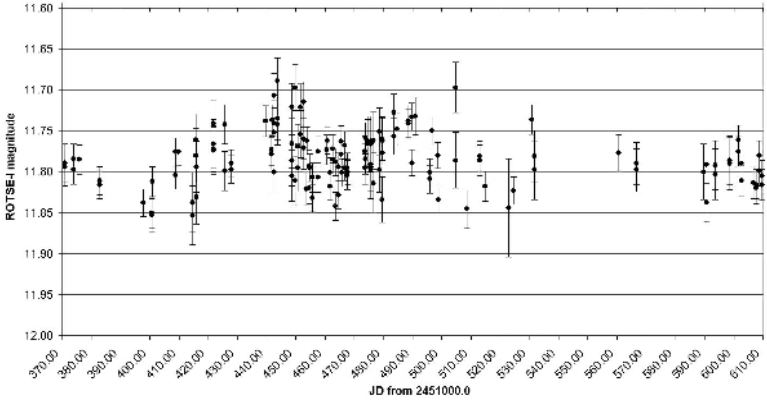

Figure 42. Light curve (NSVS dataset) of GSC 04051-02483.

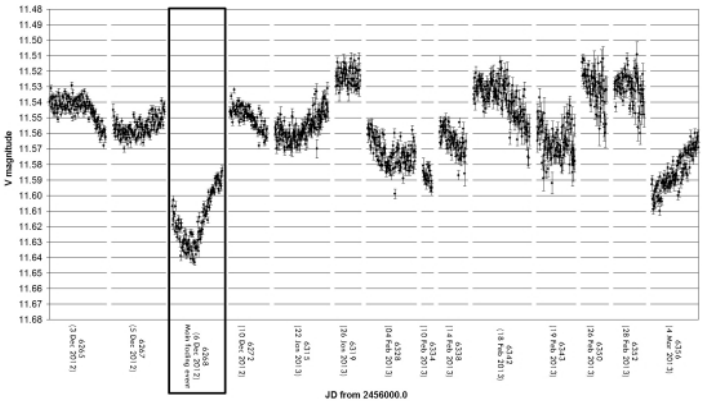

Figure 43. Light curve (Furgoni dataset) of GSC 04051-02483.

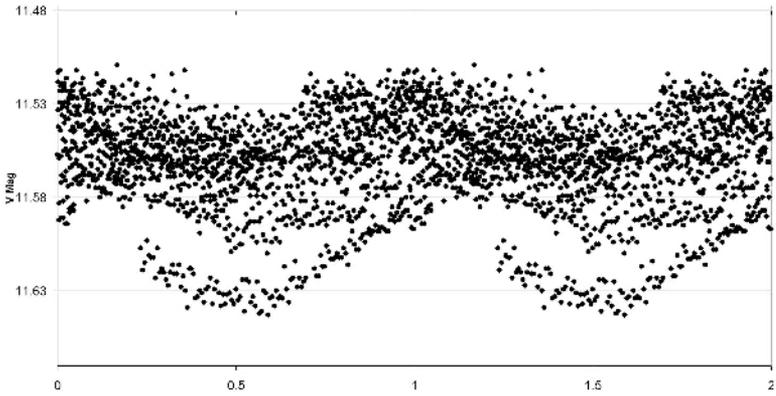

Figure 44. Phase plot of GSC 04051-02483. Furgoni dataset.

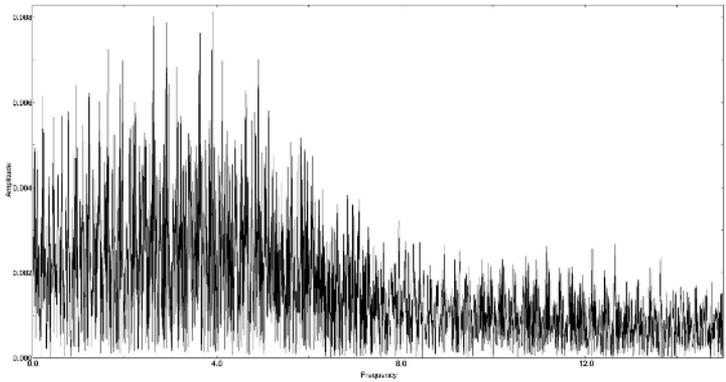

Figure 45. Fourier spectrum of GSC 04051-02483 (residuals after subtraction of frequencies (cycles per day): 0.397874538; 0.204961957; 0.834861469.

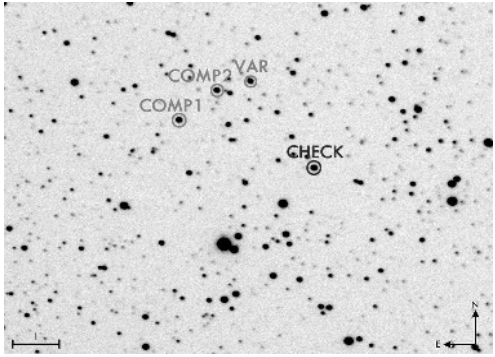

Figure 46. Finding chart of 2MASS J02472793+6149024.

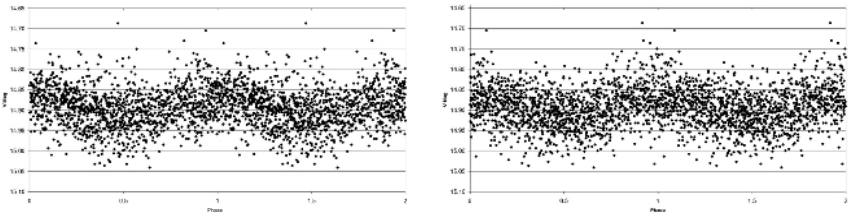

Figures 47 and 48. Main (left) and secondary (right) period phase plots of 2MASS J02472793+6149024 (Furgoni dataset).

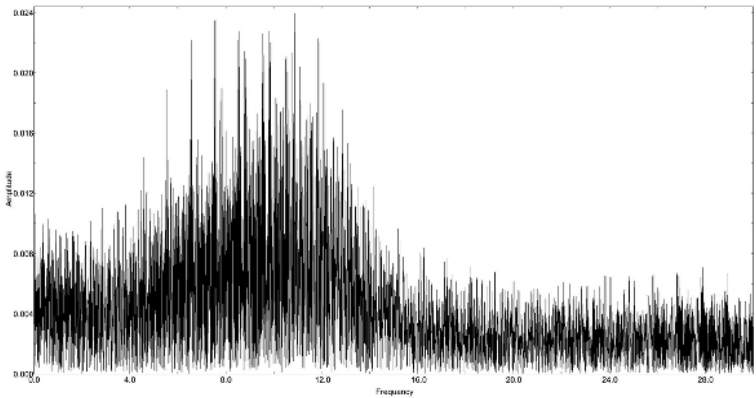

Figure 49. Fourier spectrum of 2MASS J02472793+6149024.

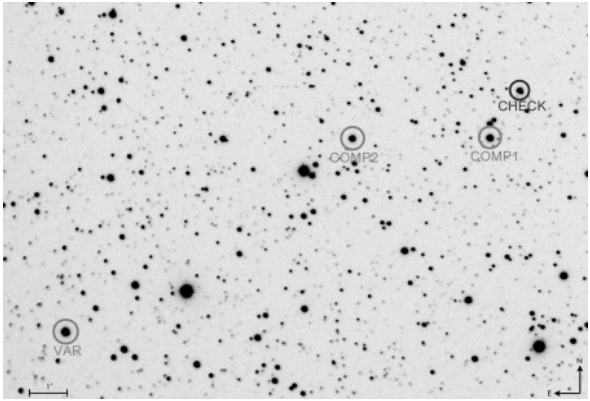

Figure 50. Finding chart of GSC 04047-01118.

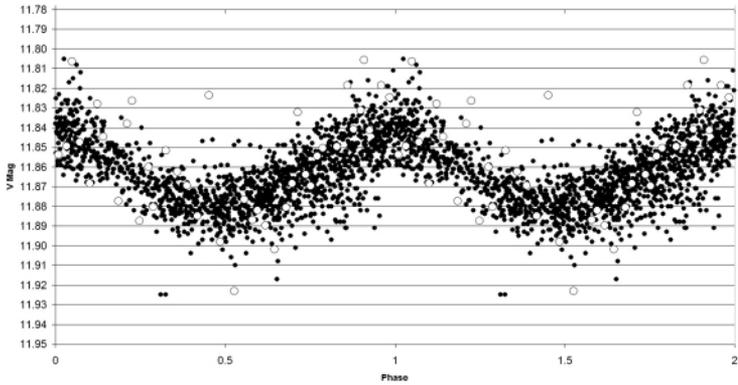

Figure 51. Phase plot of GSC 04047-01118. Filled circles denote Furgoni dataset; open circles denote NSVS dataset (–0.35 mag. offset, Binning 10).

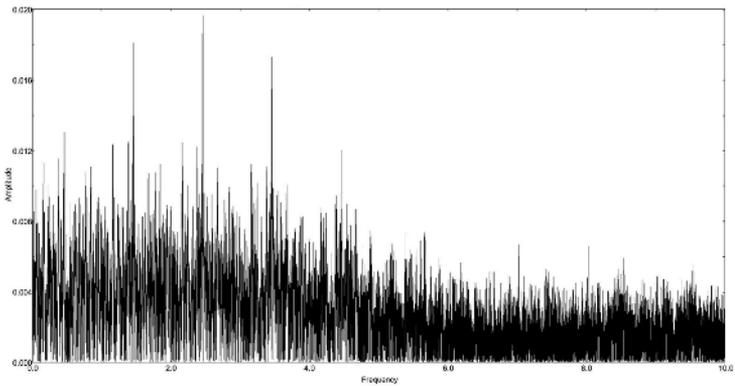

Figure 52. Fourier spectrum of GSC 04047-01118.

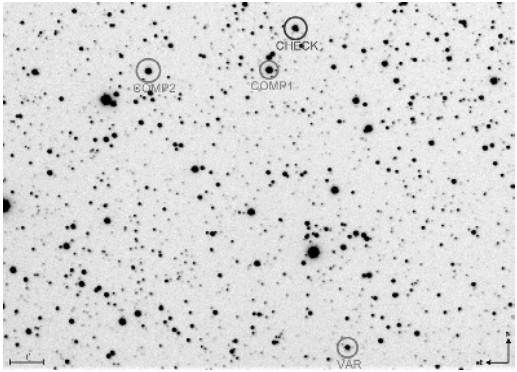

Figure 53. Finding chart of  GSC 04047-00558.

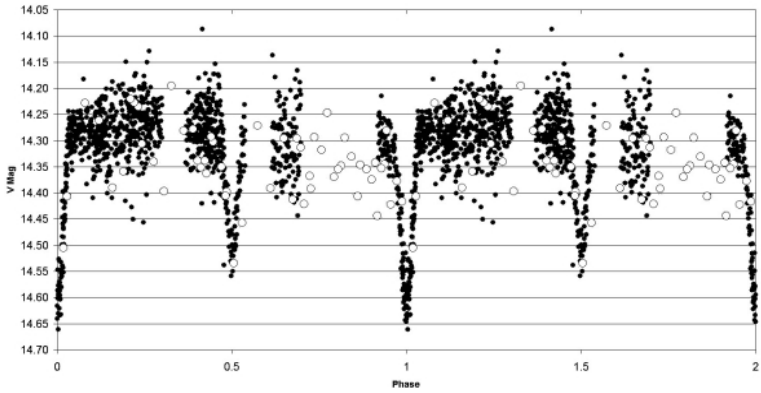

Figure 54. Phase plot of  GSC 04047-00558. Filled circles denote Furgoni dataset; open circles denote NSVS dataset (0.00 mag. offset, binning 4).

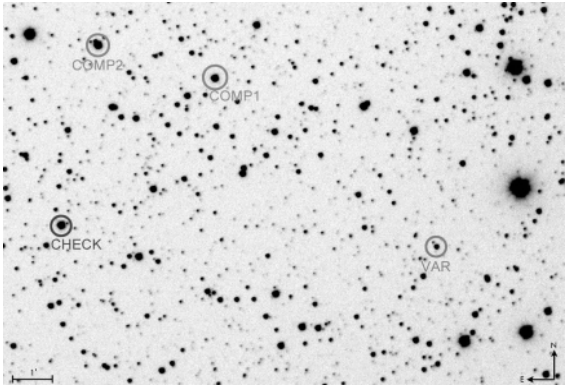

Figure 55. Light curve of GSC 04047-00381.

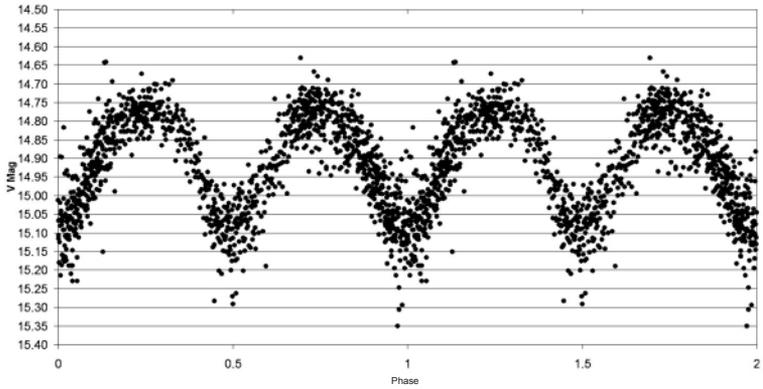

Figure 56. Phase plot of GSC 04047-00381. Furgoni dataset.

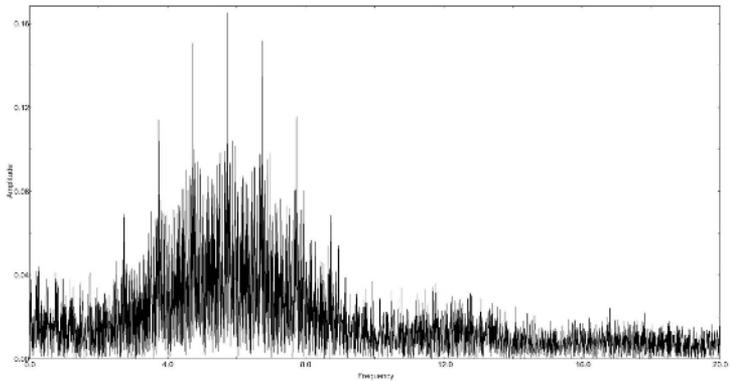

Figure 57. Fourier spectrum of GSC 04047-00381.

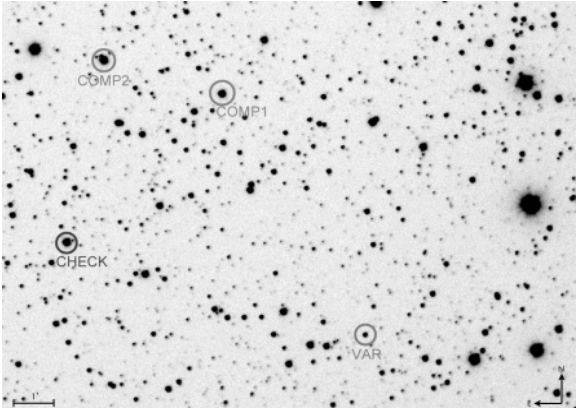

Figure 58. Finding chart of 2MASS J02443720+6143091.

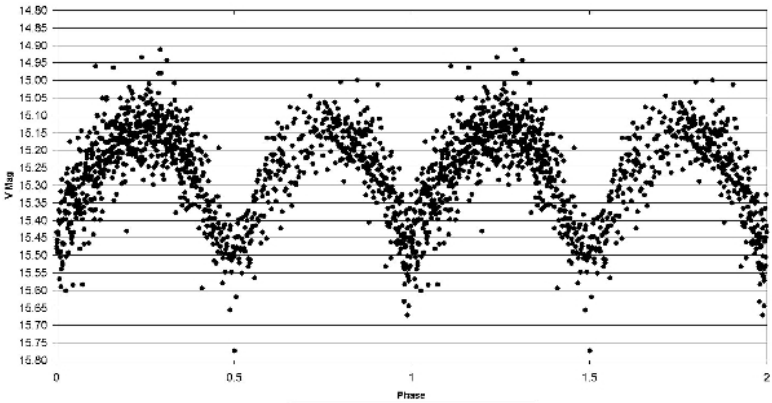

Figure 59. Phase plot of 2MASS J02443720+6143091. Furgoni dataset.

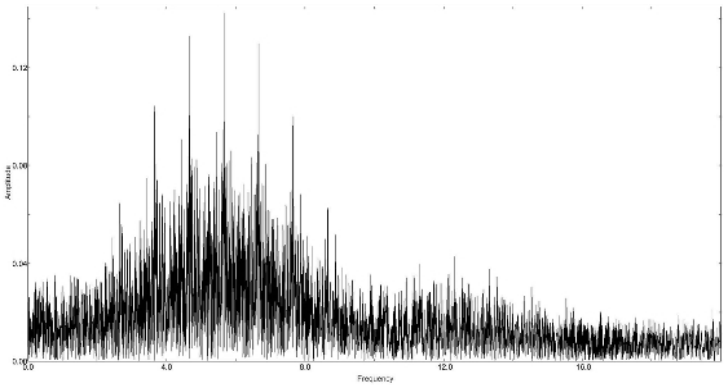

Figure 60. Fourier spectrum of 2MASS J02443720+6143091.

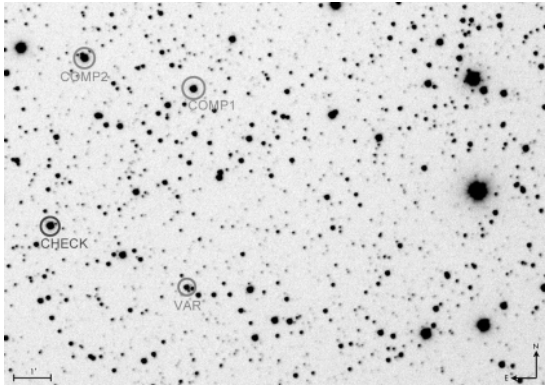

Figure 61. Finding chart of GSC 04047-01418.

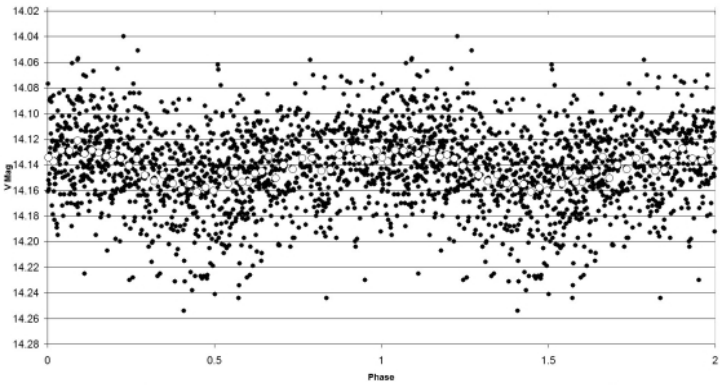

Figure 62. Phase plot of GSC 04047-01418. Filled circles denote Furgoni dataset; open circles denote Furgoni dataset, Binning 30.

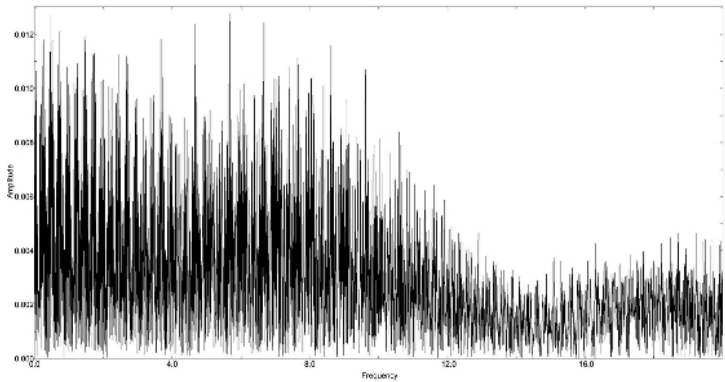

Figure 63. Fourier spectrum of GSC 04047-01418.